\documentclass[twocolumn,prl,tightenlines,superscriptaddress,showpacs]{revtex4-2}

\usepackage{amsmath}
\usepackage{amssymb,amsfonts,latexsym}
\usepackage{bm}
\usepackage[mathcal]{euscript}
\usepackage{graphicx}
\usepackage{epsfig}
\usepackage{color}

\newcommand{\be}{\begin{equation}}
\newcommand{\ee}{\end{equation}}

\newcommand{\mC}{\mathcal{C}}

\begin{document}

\title{Far-from-equilibrium complex landscapes}

\author{Laura Guislain}
\author{Eric Bertin}
\affiliation{Universit\'e Grenoble Alpes, CNRS, LIPhy, 38000 Grenoble, France}


\date{\today}

\begin{abstract}
Systems with a complex dynamics like glasses or models of biological evolution are often pictured in terms of complex landscapes, with a large number of possible collective states.
We show on the example of a stochastic spin model with non-reciprocal and heterogeneous interactions how the complex landscape notion can be generalized far from equilibrium, where collective states may exhibit spontaneous oscillations, often hidden by the presence of disorder. We identify relevant observables, like the density of entropy production, to unveil the presence of oscillations, and 
we characterize the complex landscape of our model in terms of a configurational entropy, that counts the number of nonequilibrium collective states with a given entropy production density.
\end{abstract}

\maketitle

Complex systems may be interpreted as systems made of a large number of interacting units, that can stay in many different collective states \cite{parisi2023nobel}. Such systems are often described by a rugged landscape with many minima corresponding to different possible collective states \cite{SGTFB-book}.
Examples include evolutionary dynamics in biology \cite{bank2016,visser2014}, population dynamics \cite{altieri2021},
supercooled liquids \cite{charbonneau2014,jin2017}, disordered optical media \cite{ghofraniha2015}, constraint satisfaction problems \cite{krzakala2007,mezard2005b}, or neural networks \cite{montemurro2000}.
The precise definition of the landscape may depend on the context: it may correspond, e.g., to a free energy landscape for physical systems \cite{ros2019,bussi2020}, or to a fitness landscape in the context of biological evolution \cite{sella2005}.
The existence of many collective states typically results from the presence of strong and heterogeneous interactions.
In many situations, the number of possible macroscopic state increases exponentially with system size, and one has to resort to a statistical description of those states
through the notion of configurational entropy (or complexity) that counts the number of states having a given characteristic (e.g., a given free energy or fitness) \cite{rosJPA2023,benarous2021}.

At equilibrium, like in supercooled liquids, states correspond to time-independent collective behaviors. However,
systems driven far from equilibrium by external forces or local activity may experience non-reciprocal interactions \cite{fruchart2021}, leading to a breaking of detailed balance.
The latter may result in time-dependent collective behaviors, like spontaneous temporal oscillations reported, e.g., in biochemical clocks \cite{Cao_free_energy2015,nguyen_phase_2018,Aufinger_complex2022}, populations of biological cells \cite{Kamino_fold2017,Wang_emergence2019},
assemblies of active particles \cite{saha_scalar_2020,you_nonreciprocity_2020}, population dynamics \cite{andrae_entropy_2010,Duan_Hopf2019},
or nonequilibrium spin models \cite{de_martino_oscillations_2019,daipra_oscillatory_2020,guislain2023}.
For systems displaying non-reciprocal heterogeneous interactions, the complex landscape picture may thus be extended to include time-dependent collective states.
Such a far-from-equilibrium complex landscape may be relevant to describe, e.g.,
neural network dynamics \cite{pina2018,kalitzin2014}, 
coupled biological clocks \cite{feillet2014,goldbeter2022}, 
biological evolution in changing environment \cite{das2022},
population dynamics with many interacting species \cite{ros2023}, 
glasses under cyclic shear \cite{mungan2019,szulc2022}, crumbled paper sheet \cite{shohat2022,bense2021}, or the laminar-turbulent transition \cite{yalniz2021,dauchot2012}.

In this Letter, we investigate a minimal spin model exhibiting a complex, far-from-equilibrium landscape.
Our model includes as a key ingredient non-reciprocal disordered interactions, leading to a rich phase diagram featuring multiple glassy oscillating states.
As the latter may be difficult to observe in practice due to the presence of disorder that conceals oscillations, we determine the relevant observables that explicitly exhibit oscillations.
In addition, we characterize the complex landscape in terms of a configurational entropy that counts the number of oscillating states having a given entropy production, which is a characterization of the irreversibility of the collective dynamics.

\begin{figure}[t]
    \centering
    \includegraphics[width=0.9\columnwidth]{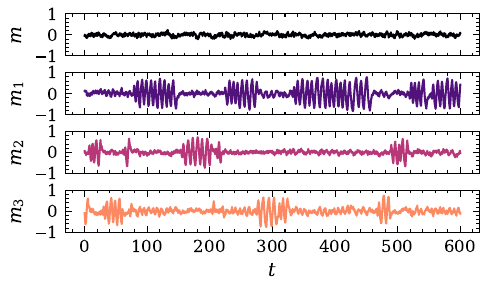}
    \caption{Magnetization $m$ and generalized magnetizations $m_{\alpha}$ versus time. Parameters: $T=0.25$, $\mu=2$, $T_g=0.4$, $T_0=0.5$, $\xi=0.2$, $N=200$ \cite{SuppMat}.}
    \label{fig:trajectories}
\end{figure}

\begin{figure}[t]
    \centering
    \includegraphics[width=0.9\columnwidth]{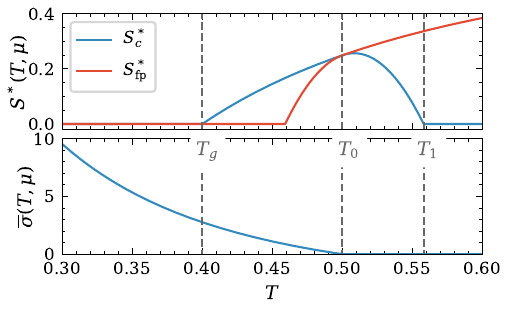}\\
    \caption{Top: Total configurational entropies $S_{c}^*(T,\mu)$ and $S_{\mathrm{fp}}^*(T,\mu)$ of oscillating and time-independent states respectively, versus temperature $T$.
    Bottom: Mean entropy production density $\overline{\sigma}(T,\mu)$ averaged over disorder [Eq.~(\ref{eq:entropy:prod:av})], versus $T$. Parameters: $\mu=2$, $T_g=0.4$, $T_0=0.5$, $\xi=0.2$.}
    \label{fig:entropy:prod}
\end{figure}

\paragraph{Generic scenario.}
Our main results regarding far-from-equilibrium complex landscape can be summarized by the following scenario, whose validity for disordered systems with non-reciprocal interactions is expected to extend beyond the specific model studied here. The numerous collective states can be classified into spontaneously oscillating states (limit cycles) and time-independent states (fixed points). The case of more complex time-dependence of collective states, like chaoticity \cite{roy2020}, is not considered here.
Importantly, oscillations in different collective states correspond to different cycles in phase space that are not visible on the same macroscopic observables.
Each oscillating state $\alpha$ may thus be characterized by its own order parameter $m_{\alpha}$, which displays oscillations only when the system is in the
collective state $\alpha$, as illustrated in Fig.~\ref{fig:trajectories}. As often the case for disordered systems, natural physical observables like magnetization display no signature of non-trivial collective states.
Oscillating states are characterized by a non-zero entropy production density $\sigma>0$ in the limit of a large system size $N$ \cite{xiao_entropy_2008,guislain2023}.
For a system with $N$ degrees of freedom described by a microscopic configuration $\mC$, and satisfying a stochastic Markov dynamics, the entropy production density is defined as
\be \label{eq:def:sigma}
\sigma = \frac{1}{N} \sum_{\mC' \ne \mC} W(\mC'|\mC) P(\mC) \ln \frac{W(\mC'|\mC)}{W(\mC|\mC')},
\ee
where $W(\mC'|\mC)$ is the transition rate from configuration $\mC$ to configuration $\mC'$, and $P(\mC)$ the probability of configuration $\mC$.
The case $\sigma=0$ corresponds to time-independent collective states (fixed points), for which the entropy production is intensive, leading to a vanishing entropy production density \cite{xiao_entropy_2008,guislain2023}.
The statistics of both oscillating and time-independent collective states can be described by densities of states $n_{c}(\sigma;\zeta)$ and $n_{\mathrm{fp}}(\zeta)$ respectively, where $\zeta$ is the set of control parameters. These densities of collective states are expected to scale exponentially with system size $N$,
which defines the corresponding configurational entropies $S_{c}(\sigma;\zeta)$ and $S_{\mathrm{fp}}^*(\zeta)$ through the relations
$n_{\mathrm{c}}(\sigma;\zeta) \sim \exp[NS_{c}(\sigma;\zeta)]$ and $n_{\mathrm{fp}}(\zeta)\sim \exp[NS_{\mathrm{fp}}^*(\zeta)]$.
Hence the entropy production density $\sigma$ plays for oscillating states a role similar to the free energy of pure states in equilibrium disordered systems at low temperature.
The total configurational entropy of oscillating states is $S_{c}^*(\zeta)=\max_{\sigma} S_{c}(\sigma;\zeta)$. We call $\sigma^*$ the value of $\sigma>0$ which maximizes $S_{c}(\sigma;\zeta)$.
When $S_{c}^*(\zeta)>S_{\mathrm{fp}}^*(\zeta)$, the macroscopic state is dominated by oscillations, and the average entropy production is $\overline{\sigma} = \sigma^*>0$,
leading to a macroscopic irreversibility. This behavior is illustrated in Fig.~\ref{fig:entropy:prod} on the explicit spin model studied below for which $\zeta=(T,\mu)$, with $T$ the bath temperature and $\mu$ a parameter quantifying the breaking of detailed balance.
In the opposite situation, $S_{c}^*(\zeta)<S_{\mathrm{fp}}^*(\zeta)$, the macroscopic state is dominated by time-independent states, as in equilibrium disordered systems,
and $\overline{\sigma}=0$ [see Fig.~\ref{fig:entropy:prod}].

\paragraph{Model.}
To illustrate this scenario on an explicit model, we study below a minimal mean-field spin model incorporating non-reciprocal interactions and slowly rearranging disordered as key ingredients.
We consider a generalization of the kinetic mean-field Ising model with ferromagnetic interactions introduced in \cite{guislain2023,guislain2024discontinuous}
(see also, e.g., \cite{collet_macroscopic_2014,collet_rhythmic_2016,Martino_feedback2019,Sinelschikov_emergence_2023} for related models).
The model involves $2N$ microscopic variables: $N$ spins $s_i=\pm1$ and $N$ fields $h_i=\pm 1$. 
We define the magnetization $m=N^{-1}\sum_{i=1}^N s_i$ and the average field $h=N^{-1}\sum_{i=1}^N h_i$.
The stochastic dynamics consists in randomly flipping a single spin $s_i$ or a single field $h_i$.
The transition rates $W_k^i$ to flip a spin $s_i=\pm1$ ($k=1$) or a field $h_i=\pm1$ ($k=2$) is given by 
\be \label{eq:trans:rate:spin:fields}
W_k^i = \frac{1}{1+\exp(\Delta_i E_{k}/T)},
\ee
with $T$ the bath temperature and $\Delta_i E_{k}$ the corresponding variation of $E_{k}$ defined as
\begin{align}\label{eq:interaction:energy:1}
E_1(\{s_i, h_i\}) &= - \frac{1}{N} \sum_{i, j}K_{ij}s_ih_j\\
\label{eq:interaction:energy:2}
E_2(\{s_i, h_i\}) &= - \frac{1}{N} \sum_{i, j}\left[(1-\mu)K_{ij}s_ih_j+\frac{J_{ij}}{2}h_ih_j\right].
\end{align}
For $\mu=0$, the transition rates (\ref{eq:trans:rate:spin:fields}) satisfy detailed balance, and the model is at equilibrium.
For $\mu \ne 0$, spins and fields have non-reciprocal interactions and detailed balance is broken.
The parameter $\mu$ can thus be interpreted as a strength of non-reciprocity, or a distance to equilibrium.
In what follows, we focus on the case $\mu>0$, as this is the regime in which oscillations may appear.
%
We choose the disorder to be separable \cite{mattis1976}: $K_{ij}=\epsilon_i\epsilon_j$ and $J_{ij}=J({\bm \epsilon})\epsilon_i\epsilon_j$,
with $\epsilon_i=\pm1$, and ${\bm \epsilon}=(\epsilon_1,\dots,\epsilon_N)$.
The amplitude $J({\bm \epsilon})$ of the ferromagnetic coupling constants is assumed to be a quenched random variable, independently drawn for each ${\bm \epsilon}$ from the distribution 
\be \label{eq:pJ}
p(J) = \sqrt{\frac{N}{\pi \xi^2}} \, \exp[-N(J-J_0)^2/\xi^2].
\ee
A given ${\bm \epsilon}$ fixes the disordered coupling constants, and thus the dynamics of the spin and field variables $s_i$ and $h_i$, leading to a statistical collective behavior interpreted as a state, that can be labeled by ${\bm \epsilon}$.
We assume a slow dynamics on ${\bm \epsilon}$ to allow the system to explore different states.
%
%
Taking inspiration from the Random Energy Model (REM) \cite{derrida.1981}, each of the $2^N$ configurations ${\bm \epsilon}$ is associated with a quenched random energy level $E({\bm \epsilon})$ drawn from the Gaussian distribution
\be
\rho(E) = \frac{1}{\sqrt{N\pi U^2}} \, \exp[-E^2/NU^2].
\ee 
The energy levels are independent random variables. The slow dynamics of ${\bm \epsilon}$ consists of randomly flipping a variable $\epsilon_i$, where $i$ is chosen at random.
We denote ${\bm \epsilon}^i$ the configuration obtained from a configuration ${\bm \epsilon}$ after flipping the variable $\epsilon_i$.
The corresponding transition rate reads
$W({\bm \epsilon}^i|{\bm \epsilon}) = \nu/[1+\exp(E({\bm \epsilon}^i)-E({\bm \epsilon}))/T]$.
We assume $\nu\ll 1$, so that the dynamics of the disorder configuration ${\bm \epsilon}$ is much slower than the dynamics of the spins $s_i$ and fields $h_i$,
and can be considered as quasistatic.
As the dynamics of ${\bm \epsilon}$ does not depend on the spins $s_i$, the fields $h_i$ and on the coupling constant $J({\bm \epsilon})$,
it satisfies detailed balance with respect to the equilibrium distribution
$P_{\mathrm{eq}}({\bm \epsilon}) \propto \exp[-E({\bm \epsilon})/T]$.
We relabel the $2^N$ energy levels $E({\bm \epsilon})$ into an increasing sequence $E_1<E_2<\dots<E_{2^N}$, and we call $\{w_{\alpha}\}_{1\leq \alpha\leq 2^N}$
the probability weights 
\be \label{eq:wa}
w_{\alpha} = \frac{\exp(- E_{\alpha}/T)}{\sum_{\alpha'} \exp(- E_{\alpha'}/T)}.
\ee
We denote ${\bm \epsilon}^{\alpha}=(\epsilon_1^{\alpha}, ...,\epsilon_N^{\alpha})$ and $J_{\alpha}=J({\bm \epsilon}^{\alpha})$ the configuration and the coupling constant associated with configuration $\alpha$. 
Similarly to the REM, the glass transition temperature is given by $T_g = U/(2\sqrt{\ln2})$ \cite{derrida.1981}.
For $T>T_g$, all configurations have comparable probability weights $w_{\alpha}$, whereas for $T<T_g$, only the few lowest energies levels are explored.

Introducing the state-dependent variables $s_i^{\alpha}=\epsilon_i^{\alpha}s_i$ and $h_i^{\alpha}=\epsilon_i^{\alpha}h_i$, 
the dynamics of $s_i^{\alpha}$ and $h_i^{\alpha}$ is the same as the dynamics of $s_i$ and $h_i$ in the absence of disorder, with a coupling constant
$J_{\alpha}$ \cite{SuppMat}. In the quasistatic limit, the joint steady-state probability distribution $P_N(\{s_i, h_i\})$ can thus be decomposed over states $\alpha$ as:
\be \label{eq:PN:s1h1sNhN}
P_N(\{s_i, h_i\}) = \sum_{\alpha} w_{\alpha}P_N^0(\{s_i^{\alpha}, h_i^{\alpha}\}; J_{\alpha}),
\ee
where $P_N^0(\{s_i, h_i\}; J)$ is the probability to have the configuration $\{s_i, h_i\}$ with coupling constant $J$ in the model without disorder (i.e., with all $\epsilon_i=+1$).
Note that the time-dependent distribution $P_N(\{s_i, h_i\},t)$ converges to a stationary distribution even in the presence of oscillating states, in which case it describes an ensemble of oscillating systems
with uniformly distributed phases \cite{guislain2023}.

\begin{figure}[t]
    \centering
    \includegraphics[width=0.9\columnwidth]{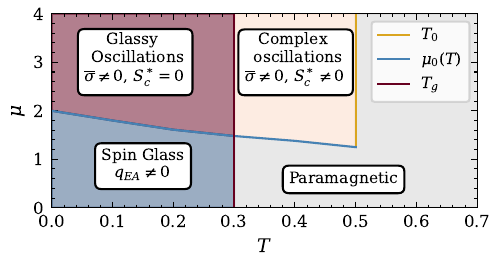}
    \caption{Schematic phase diagram for $T_g<T_0$ in the presence of disorder. Macroscopic oscillations are found for $T<T_0$ and $\mu>\mu_0(T)$.
    The configurational entropy $S_c^*$ of oscillating states is nonzero for $T>T_g$.}
    \label{fig:phase:diagram}
\end{figure}

\paragraph{Phase diagram.}
For a high enough temperature $T$, the model is in a paramagnetic phase for all values of $\mu$.
In contrast, the low-temperature behavior of the model depends on the non-reciprocity parameter $\mu$: for $\mu<\mu_0(T)$, the model is in a spin-glass phase, while for $\mu>\mu_0(T)$
a phase with multiple oscillating states is found (see Fig.~\ref{fig:phase:diagram}).
The spin-glass phase can be detected by a nonzero value of the Edwards-Anderson parameter 
$q_{EA} = N^{-1} \sum_i \overline{\langle s_i \rangle^2}$ \cite{mezard_spin_1987}.
In the infinite $N$ limit, we find 
that $q_{EA}$ is non-zero only for $T<T_g$ and $\mu<\mu_0(T)$ \cite{SuppMat},
indicating a spin-glass phase in this region of the phase diagram (Fig.~\ref{fig:phase:diagram}).
Oscillations present at low temperature for $\mu>\mu_0(T)$ are hidden by disorder and are not visible on the magnetization $m(t)$ (top panel of Fig.~\ref{fig:trajectories}).
Yet, they can be detected by evaluating the entropy production (see below).
They can also be visualized by introducing the state-dependent observables
\be \label{eq:change:var:mh}
m_{\alpha}=N^{-1} \sum_i s_i^{\alpha}, \qquad h_{\alpha}=N^{-1}\sum_i h_i^{\alpha}.
\ee
The generalized magnetization $m_{\alpha}(t)$ is plotted in Fig.~\ref{fig:trajectories} for $\alpha=1$, $2$ and $3$, corresponding to the three lowest 
energy levels $E_{\alpha}$ (see \cite{SuppMat} for the simulation method).
Oscillations are visible for each of these observables over some limited time windows, corresponding to the time spent in state $\alpha$.

\paragraph{Entropy production.}
The onset of spontaneous oscillations can be quantified by the entropy production density, whose nonzero value in the limit $N\to \infty$ indicates
macroscopic irreversibility.
We define the entropy production density $\sigma_{\alpha}(T,\mu)$ in state $\alpha$,
\be
\sigma_{\alpha}(T,\mu) = \frac{1}{N} \sum_{\mC' \ne \mC} W_{\alpha}(\mC'|\mC) P_{\alpha}(\mC) \ln \frac{W_{\alpha}(\mC'|\mC)}{W_{\alpha}(\mC|\mC')}
\ee 
where $\mC$ corresponds to a configuration $\mC=\{s_i, h_i\}$.
We find $\sigma_{\alpha}(T,\mu) = \sigma(J_{\alpha};T,\mu)$, with for $J_{\alpha}>2T$
\be \label{eq:sigma:Jalpha}
\sigma(J_{\alpha};T,\mu) = A(T,\mu)\frac{J_{\alpha}-2T}{J_{\alpha}},
\ee
to first order in an expansion in $J_{\alpha}-2T$, 
and $\sigma(J_{\alpha};T,\mu)=0$ otherwise (see \cite{SuppMat} for the expression
of $A(T,\mu)$).
Averaging over states $\alpha$ and over the disorder, we obtain
$\overline{\sigma}(T,\mu) = \overline{\sum_{\alpha} w_{\alpha} \sigma(J_{\alpha};T,\mu)}$
where the overbar stands for an average over the disorder $(E_{\alpha},J_{\alpha})$.
As $w_{\alpha}$ depends only on $E_{\alpha}$, we get
\be \label{eq:entropy:prod:av}
\overline{\sigma}(T,\mu) = \int dJ\, p(J)\, \sigma(J;T,\mu).
\ee
The average entropy production density $\overline{\sigma}(T,\mu)$ is plotted in Fig.~\ref{fig:entropy:prod} (bottom panel),
in the limit $N\to \infty$.
One finds that $\overline{\sigma}(T,\mu)$ is non zero only for $T<T_0 = J_0/2$ and $\mu>\mu_0(T)$. 
This confirms the presence of spontaneous oscillations in this parameter range.

\begin{figure}[t]
    \centering
    \includegraphics[width=0.9\columnwidth]{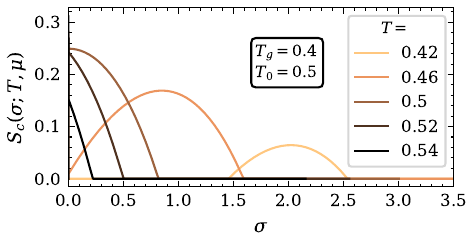}
    \caption{Configurational entropy $S_c(\sigma;T,\mu)$ as a function of the entropy production density $\sigma$, for different temperatures. Parameters: $\mu=2$, $T_g=0.4$, $T_0=0.5$, $\xi=0.2$. }
    \label{fig:entropy_config}
\end{figure}

\paragraph{Configurational entropy of oscillating states.}
We now assume $T_g<T_0$.
In the case $T<T_g$ and $\mu>\mu_0(T)$ illustrated in Fig.~\ref{fig:trajectories}, only a few oscillating states $\alpha$ are present, the ones corresponding to the lowest energies $E_{\alpha}$.
For $T>T_g$, the number of oscillating states 
grows exponentially with system size, $\mathcal{N}^*(T,\mu) \sim \exp[N S_c^*(T,\mu)]$, where $S_c^*(T,\mu) \in (0,\ln 2)$ is the global configurational entropy of oscillating states.
A finer characterization is given by considering the number of states with a given entropy production density, $\mathcal{N}(\sigma;T,\mu) \sim \exp[NS_c(\sigma; T,\mu)]$.
The statistics of $\sigma$ can be obtained from the statistics of the coupling $J$, by inverting Eq.~(\ref{eq:sigma:Jalpha}) for $J_{\alpha}>2T$ to get $J(\sigma;T,\mu)$.
We introduce the density of state $n(e,\sigma;T,\mu)$, with $e=E/N$ the energy density.
The average density of states over the disorder is $\overline{n}(e,\sigma; T,\mu) = 2^N N\rho(Ne) p(J(\sigma;T,\mu)) \sim \exp\left[N s(e, \sigma;T,\mu)\right]$ with 
\be
s(e, \sigma; T,\mu) = \ln2-\frac{e^2}{U^2} - \frac{1}{\xi^2} \big(J(\sigma; T,\mu)-J_0\big)^2.
\ee
When the density of states $n(e,\sigma; T,\mu)$ is large, its relative fluctuations are small and $n(e, \sigma; T,\mu) \sim \overline{n}(e, \sigma;T,\mu)$. 
If $s(e, \sigma; T,\mu)>0$, corresponding to $|e|<e_0$ with
\be
e_0 = U\sqrt{\ln 2-\big(J(\sigma; T,\mu)-J_0\big)^2/\xi^2},
\ee
$\overline{n}(e, \sigma; T,\mu)$ is exponentially large with $N$ so that the assumption $\overline{n}(e, \sigma; T,\mu) \sim n(e, \sigma; T,\mu)$ is valid. However for $|e|>e_0$,
$\overline{n}(e, \sigma; T,\mu)$ is exponentially small with $N$, which means that in most samples, there are no configurations at energy density $|e|>e_0$. 
Energy states are reweighted by a Boltzmann factor, so that the most probable energy density $e^*$ satisfies
$\partial_e s(e, \sigma; T,\mu)=1/T$, yielding $e^*=-U^2/2T$, as long as $e^*>-e_0$ (otherwise, $e^*=-e_0$, see \cite{SuppMat}).
We thus get for the configurational entropy $S_c(\sigma; T,\mu)=s(e^*,\sigma;T,\mu)$, leading to
\be \label{eq:Sc:sigma}
S_c(\sigma; T,\mu) = \ln 2-\frac{1}{\xi^2} \big(J(\sigma; T,\mu)-J_0\big)^2-\frac{U^2}{4T^2}
\ee
if $\sigma_{\min}(T,\mu) < \sigma < \sigma_{\max}(T,\mu)$, and $S_c(\sigma; T,\mu)=0$ otherwise.
The bounds $\sigma_{\min}$ and $\sigma_{\max}$ are the two values of $\sigma$ for which the expression of $S_c(\sigma; T,\mu)$ given in Eq.~(\ref{eq:Sc:sigma}) vanishes.
%
For $T<T_g$, $S_c(\sigma;T,\mu)=0$ for all $\sigma$.
The configurational entropy $S_c(\sigma; T,\mu)$ is plotted in Fig.~\ref{fig:entropy_config}.
We define the total configurational entropy (see Fig.~\ref{fig:entropy:prod})
\be
S_c^{*}(T, \mu)=\underset{{\sigma>0}}{\max} \,S_c(\sigma; T, \mu).
\ee
We find that $S_c^{*}(T,\mu)$ is nonzero only over a finite temperature range $(T_g, T_1)$, where $T_1>T_0$ \cite{SuppMat}. 
For $T_g<T< T_1$, $S_c^{*}(T)>0$, leading to an exponentially large number of oscillating states (Fig.~\ref{fig:entropy:prod}).
However, to know whether oscillating states dominate over time-independent states (fixed points), one needs to compare their configurational entropies.
The configurational entropy $S_{\mathrm{fp}}^*(T,\mu)$ of fixed points, i.e., states with an entropy production density $\sigma=0$, can be evaluated along similar lines
\cite{SuppMat}, and is plotted in Fig.~\ref{fig:entropy:prod}.
For $T>T_0$, $S_{\mathrm{fp}}^*(T,\mu)>S_c^*(T,\mu)$ and time-independent states are exponentially more numerous than oscillating states, which are thus in practice not observed for large system sizes.
Accordingly, the mean entropy production density $\overline{\sigma}$ vanishes (Fig.~\ref{fig:entropy:prod}).
In contrast, for $T_g < T < T_0$, $S_{\mathrm{fp}}^*(T,\mu)<S_c^*(T,\mu)$ and oscillating states dominate over time-independent states. The macroscopically observed state is thus genuinely out of equilibrium,
with $\overline{\sigma}>0$, indicating macroscopic irreversibility.

\paragraph{Conclusion.}
We have studied in a disordered spin model with non-reciprocal couplings the emergence of a far-from-equilibrium complex landscape, in which multiple collective oscillating states may exist.
These collective states are non-trivial in the sense that oscillations in different states cannot be observed on the same observable (Fig.~\ref{fig:trajectories}).
In some parameter regimes, the number of oscillating states grows exponentially with system size, and can be quantified by a configurational entropy.
This description can be refined by counting the number of oscillating states with a given entropy production density.
Oscillating states may statistically compete with time-independent states, and their respective configurational entropies determine whether oscillating or time-independent states dominate on average.
The notion of a far-from-equilibrium complex landscape is expected to be relevant in many different contexts, for systems exhibiting disordered and non-reciprocal interactions, ranging from
fitness models in biological evolution to neural networks, coupled biological clocks, or models of heterogeneous socio-economic agents.
Future work might try to enrich further the far-from-equilibrium landscape picture by also including chaotic states \cite{roy2020}.

\newpage 

\begin{widetext} 
  \begin{center}\textbf{\large SUPPLEMENTAL MATERIAL: Far-from-equilibrium complex landscapes}\end{center}
\end{widetext}

\setcounter{equation}{0}
\setcounter{figure}{0}

\section{Phase diagram in the absence of disorder}
\subsection{State-dependent variables}
From a configuration $\bm \epsilon^{\alpha}=(\epsilon_1^{\alpha}, ..., \epsilon_N^{\alpha})$, we introduce the state-dependent variables $s_i^{\alpha}=\epsilon_i^{\alpha}s_i$ and $h_i^{\alpha}=\epsilon_i^{\alpha}h_i$. The energies involved in the transition rates introduced in the main text Eqs.~(3) and (4) can be rewritten as 
\begin{align}\label{eq:interaction:energy:1:changevar}
E_1(\{s_i, h_i\}) &= - \frac{1}{N} \sum_{i, j}s_i^{\alpha}h_j^{\alpha}\\
\label{eq:interaction:energy:2:changevar}
E_2(\{s_i, h_i\}) &= - \frac{1}{N} \sum_{i, j}\left[(1-\mu)s_i^{\alpha}h_j^{\alpha}+\frac{J_{\alpha}}{2}h_i^{\alpha}h_j^{\alpha}\right],
\end{align}
corresponding to the energies of the model with $K_{ij}=K$ and $J_{ij}=J_{\alpha}$. 
Thus, the dynamics of $s_i^{\alpha}$ and $h_i^{\alpha}$, while the system is in a configuration $\bm \epsilon^{\alpha}$, is the same as the dynamics of $s_i$ and $h_i$ in the absence of disorder, with a coupling constant $J_{\alpha}$.

\subsection{Phase diagram of the model without disorder}
\begin{figure}[t]
    \centering
    \includegraphics[width=\columnwidth]{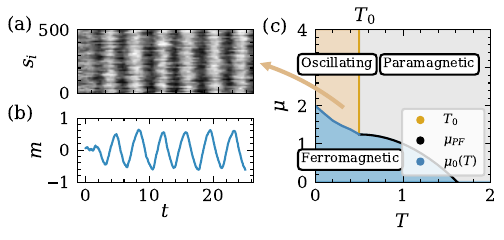}
    \caption{Schematic phase diagram for $T_g<T_0$ without disorder.}
    \label{fig:phase:diagram:nodisorder}
\end{figure}
In the absence of disorder, i.e., for $K_{ij}=K$ and $J_{ij}=J$, the model exhibits three phases, as sketched in Fig.~\ref{fig:phase:diagram:nodisorder}(c):
a paramagnetic phase at high temperature $T$, a ferromagnetic phase at low $T$ and low $\mu$, and an oscillating phase where the magnetization spontaneously oscillates in time at low $T$ and high $\mu$ [Fig.~\ref{fig:phase:diagram:nodisorder}(a,b)] \cite{guislain2023_SM}.
We call $\mu_{PF}(T)$ the transition line between paramagnetic and ferromagnetic states, $\mu_{0}(T)$ the transition line between oscillating and ferromagnetic states.
The transition line between oscillating and paramagnetic states corresponds to $T=T_0\equiv J/2$ and $\mu_{PF}(T_0)$.

\section{Edwards-Anderson parameter}
The presence of a spin glass phase can be detected by a non-zero value of the Edwards-Anderson parameter \cite{mezard_spin_1987_SM}:
\be
q_{EA} = \frac{1}{N} \sum_{i=1}^N \overline{\langle s_i \rangle^2}.
\end{equation} 
Using Eq~(8) of the main text, we obtain
\begin{equation} \langle s_i\rangle=\sum_{\alpha}w_{\alpha}\epsilon_i^{\alpha}\sum_{\{s_i^{\alpha}, h_i^{\alpha}\}}s_i^{\alpha}P^0(\{s_i^{\alpha}, h_i^{\alpha}\}; J_{\alpha}).\end{equation}
The mean magnetisation in the absence of disorder for a coupling constant $J$ is introduced as
\be m_0(J)=\sum_{\{s_i^{\alpha}, h_i^{\alpha}\}}s_i^{\alpha}P^0(\{s_i^{\alpha}, h_i^{\alpha}\}; J).\ee
We then obtain
\begin{equation} q_{EA}=\frac{1}{N} \sum_{i=1}^N \overline{\left(\sum_{\alpha}w_{\alpha}\epsilon_i^{\alpha}m_0(J_{\alpha})\right)^2}\,.\end{equation} 
As $w_{\alpha}$ depends only on $E_{\alpha}$, and $J_{\alpha}$ and $\bm\epsilon$ are chosen independently, the average over $w_{\alpha}$, $\bm\epsilon$ and $J_{\alpha}$ can be factorised. The average $\epsilon_i^{\alpha}\epsilon_i^{\beta}$ is zero when $\alpha\neq\beta$. 
We can therefore rewrite the Edwards-Anderson order parameter as follows 
\begin{equation} \label{GLASSY:eq:qEA:factorized}
q_{EA} = \left( \sum_{\alpha} w_{\alpha}^2 \right) \int dJ\, p(J)\, m_0(J)^2.
\end{equation} 
In the limit $N\to\infty$, we find that $\sum_{\alpha} w_{\alpha}^2=1-T/T_g$ for $T<T_g$ (and zero otherwise), while the magnetisation $m_0(J)$ is only nonzero for $\mu<\mu_{0}$.
Eq.~(\ref{GLASSY:eq:qEA:factorized}) shows that $q_{EA}$ is nonzero only for $T<T_g$ and $\mu<\mu_{0}$, which indicates the presence of a spin glass phase in this region of the phase diagram (see Fig~3 of the main text).

\section{Stochastic simulations of the model}
The stochastic model introduced in the main text contains $2^N$ configurations $\bm \epsilon=(\epsilon_1, ..., \epsilon_N)$, and each configuration is associated with an energy level. 
To run simulations of the model, one needs to calculate the $2^N$ energy levels (drawn at random) and determine the associated configurations $\bm \epsilon$. Numerically, the size $N$ of the system is quite limited; for example, for $N=100$ alone, one would need to calculate more than $10^{30}$ configurations. 

We propose an alternative method for carrying out simulations below the critical temperature $T_g$.
Below the critical temperature, the $w_{\alpha}$ weights associated with the $\bm \epsilon$ configurations follow a Poisson-Dirichlet law, 
and the configuration weights concentrate on a finite number of configurations. 
We draw the $n$ ($n\ll 2^N$) first $w_{\alpha}$ weights according to the Poisson-Dirichlet law of parameter $T/T_g$ (the $w_{\alpha}$ weights are on average ordered in a decreasing manner).
For each weight $w_{\alpha}$ associated with a configuration $\alpha$, we randomly draw the variables $\epsilon_i^{\alpha}=\pm 1$ and a value of $J_\alpha$ which is selected in accordance with Eq.~(5) of the main text.

According to \cite{arous.etal.2002_SM}, the long-time dynamics of the REM can be mapped to a trap model dynamics where each state $\alpha$ is associated with an average escape time $\tau_{\alpha}\equiv w_{\alpha}/\lambda$, over time scales that are exponential in $N$, where $N$ is the number of spins in the REM.
We use this mapping to simulate efficiently the long-time dynamics of our model in rescaled time units, as shown in Fig.~1 of the main text.
Thus, the dynamics of the configurations $\bm \epsilon$ (and thus of the states $\alpha$) are chosen as follows: a state $\alpha$ among the $n$ possible is chosen randomly and associated with a time $\tau$ exponentially distributed according to 
\be \label{eq:tau:law}
p_{\alpha}(\tau) =\tau_{\alpha}^{-1} e^{-\tau/\tau_{\alpha}}.
\ee
After this time $\tau$, a new state $\alpha'$ is again chosen randomly and a new time $\tau$ is again drawn according to Eq.~(\ref{eq:tau:law}) with a characteristic time $\tau_{\alpha'}$. 
We consider that the dynamics of configurations $\bm \epsilon$  is much slower than that of spins and fields, which corresponds to $\lambda\ll 1$ as $\lambda$ is proportional to the attempt frequency $\nu$ of the microscopic dynamics, introduced in the main text. Thus, during each time interval $\tau$, we maintain the same spin and field reversal dynamics as introduced in the main text.
For Fig.~1 of the main text, we take $\lambda=10^{-2}$ and $n=100$.

\section{Entropy production}
For a system with $N$ degrees of freedom described by a microscopic configuration $\mC$, and satisfying a stochastic Markov dynamics, the entropy production density is defined as
\be \label{eq:def:sigmaSM}
\sigma = \frac{1}{N} \sum_{\mC'\neq\mC} W(\mC'|\mC) P_N(\mC) \ln \frac{W(\mC'|\mC)}{W(\mC|\mC')},
\ee
where $W(\mC'|\mC)$ is the transition rate from configuration $\mC$ to configuration $\mC'$, and $P(\mC)$ the probability of configuration $\mC$.

Since the distribution of configurations $\bm\epsilon$ verifies detailed balance, it does not contribute to the entropy production. Using Eq.~(8) of the main text, we thus find 
\be \label{eq:def:sigmaSM2}
\sigma = \sum_{\alpha}w_{\alpha}\frac{1}{N} \sum_{\mC'\ne \mC} W(\mC'|\mC) P_N^0(\mC^{\alpha}; J_{\alpha}) \ln \frac{W(\mC'|\mC)}{W(\mC|\mC')},
\ee
where $\mC=\{s_i, h_i\}$ and $\mC^{\alpha}=\{s_i^{\alpha}, h_i^{\alpha}\}$.
Writing the transition rates $W(\mC'|\mC)$ in the variables $s_i^{\alpha}$ and $h_i^{\alpha}$, we introduce
$W(\mC'|\mC)=W_{\alpha}({\mC'}^{\alpha}|\mC^{\alpha})$. From its definition, one has that $W_{\alpha}({\mC'}|\mC)$ corresponds to the transition rates of the model without disorder with a coupling constant $J_{\alpha}$. Finally, we have
\be \label{eq:def:sigmaSM3}
\sigma = \sum_{\alpha}w_{\alpha}\sigma_{\alpha},
\ee
where 
\be \sigma_{\alpha}=\frac{1}{N} \sum_{\mC' \ne \mC} W_{\alpha}({\mC'}^{\alpha}|\mC^{\alpha}) P_N^0(\mC^{\alpha}; J_{\alpha}) \ln \frac{W_{\alpha}({\mC'}^{\alpha}|\mC^{\alpha})}{W_{\alpha}(\mC^{\alpha}|{\mC'}^{\alpha})}.\ee
As the change of variables from $s_i$ to $s_i^{\alpha}$ and from $h_i$ to $h_i^{\alpha}$ is linear with a Jacobian of $1$, one  finds 
\be \sigma_{\alpha}=\frac{1}{N} \sum_{\mC'\ne \mC} W_{\alpha}({\mC'}|\mC) P_{\alpha}(\mC) \ln \frac{W_{\alpha}({\mC'}|\mC)}{W_{\alpha}(\mC|{\mC'})},\ee
where $P_{\alpha}(\mC)=P_N^0(\mC; J_{\alpha})$. As $W_{\alpha}(\mC'|\mC)$ and $P_{\alpha}(\mC)$ correspond to the transition rates and the probability density of the model without disorder with a coupling constant $J_{\alpha}$, $\sigma_{\alpha}(T,\mu)$ is the entropy production density of the model without disorder with a coupling constant $J_{\alpha}$. We write $ \sigma_{\alpha}(T,\mu)= \sigma(J_{\alpha};T,\mu)$.

Using a large $N$ and small $m=N^{-1}\sum_i s_i$ and $h=N^{-1}\sum_i h_i$ approximations (valid for $J_{\alpha}\sim 2T$), one has for $\mu>\mu_0(T)$ \cite{guislain2024discontinuous_SM} 
\be \sigma_{\alpha}=\frac{(1+2T^2)(\mu-1-J_{\alpha}T-J_{\alpha}^2)}{T^2}\langle m^2\rangle\ee
where $\langle.\rangle$ is the statistical average. Using a large deviation approximation \cite{guislain2023_SM}, one finds, for $\mu>\mu_0(T)$ and at the lowest order in $J_{\alpha}-2T$,
\be \langle m^2\rangle=\frac{\mu}{(\mu-1)^2}\frac{J_{\alpha}-2T}{J_{\alpha}}.\ee 
Finally, at the lowest order in $J_{\alpha}-2T$, we have  \begin{equation} \label{GLASSY:eq:sigma:JalphaSM}\sigma(J_{\alpha},T) =A( T,\mu)\frac{J_{\alpha}-2T}{J_{\alpha}}\theta(J_{\alpha}-2T)
\end{equation} 
where $\theta(x)$ is the Heaviside function that satisfies $\theta(x)=1$ for $x>0$ and $\theta(x)=0$ for $x<0$ and 
\begin{equation} A(T,\mu)=\frac{(1+2T^2)\mu(\mu-1-T^2)}{(\mu-1)^2T^2}\end{equation} 
for $\mu>\mu_0(T)$. For $\mu<\mu_0(T)$, $\sigma(J_{\alpha}, T)=0$.

\section{Configurational entropy of oscillating states}
\subsection{Derivation of the configurational entropy}
Since the distribution of configurations $\bm\epsilon$ verifies detailed balance with respect to the equilibrium distribution $P_{\mathrm{eq}}(\bm\epsilon)\propto \exp[-E(\bm\epsilon)/T]$, we can evaluate the partition function of the configurations $\bm\epsilon$ :
\begin{equation} Z(\sigma ; T, \mu)=\sum_{\alpha=1}^{2^N}e^{-E_{\alpha}/T},\end{equation}
which can be rewritten as an integral over the energy density $e$:
\begin{equation} Z(\sigma; T, \mu)=\int_{-\infty}^{\infty}de\, n(e, \sigma; T, \mu)e^{-Ne/T}\,,\end{equation}
with $n(e, \sigma; T, \mu)$ the density of states.

In the main text, we show that for $|e|<e_0$ the assumption $\overline{n}(e,\sigma; T, \mu)\sim n(e,\sigma; T, \mu)$ is valid, with $\overline{n}(e,\sigma; T,\mu) \sim \exp\left[N s(e, \sigma;T,\mu)\right]$ and $s(e, \sigma;T,\mu)$ is given in Eq.~(13) of the main text. For $|e|>e_0$, $\overline{n}(E,\sigma; T, \mu)$ is exponentially small with $N$, which means that in most samples there are no configurations at energy density $|e|>e_0$.  
Thus, the typical value of $Z$ can be estimated as 
\begin{equation} Z_{\mathrm{typ}}(\sigma;T,\mu)\sim \int_{-e_0}^{e_0}de\, \exp\left[N\left(s(e, \sigma; T,\mu)-e/T\right)\right]\,.\end{equation}
In the limit of large $N$, we can evaluate $Z_{\mathrm{typ}}$ by a saddle-node approximation. 
We denote $e^*(T)$ the value of $e$ achieving the maximum of $s(e,\sigma; T,\mu)-e/T$ if it is in $[-e_0, e_0]$ and $e^*=-e_0$ otherwise.
We find $e^*=-U^2/2T$ if $e^*>-e_0$. 
Finally, we denote $S_c(\sigma ; T,\mu)=s(e^*(T), \sigma; T, \mu)$, yielding Eq.~(15) of the main text.
Taking the maximum of $S_c(\sigma ; T,\mu)$ over $\sigma$ leads to the total configurational entropy $S_c^*(T,\mu)$. 

\subsection{Discussion on the value of $S_{c}^*(T, \mu)$}
We now discuss the value of $S_{c}^*(T, \mu)$ for $\mu>\mu_0(T)$. 
For $T<T_g$, $S_c(\sigma ; T,\mu)=0$ thus $S_c^*(T,\mu)=0$. 
According to the main text, $\sigma(J; T, \mu)>0$ for all $T\leq J/2$. 
Thus, for $T_g\leq T\leq T_0$, where $T_0\equiv J_0/2$, the maximum is reached in $J(\sigma; T,\mu)=J_0$ which leads to 
\be S_c^*(T, \mu)=\ln 2-\frac{U^2}{4T^2}.\ee
For $T>T_0$, the maximum of $S_c(\sigma ; T,\mu)$ is reached when $\sigma(J; T, \mu)\to 0$ corresponding to $J\to 2T$. Thus, 
\be S_c^*(T,\mu)=\ln 2-\frac{(2T-J_0)^2}{\xi^2}-\frac{U^2}{4T^2}\ee
for $T_0<T<T_1$ where $T_1>T_0$ is one solution of $\ln 2 -(2T-J_0)^2/\xi^2-U^2/(4T^2)=0$. For $T>T_1$, $ S_c^*(T,\mu)=0$. $S_{c}^*(T, \mu)$ is plotted in Fig.~2 of the main text. 

\medskip

\section{Configurational entropy of fixed points} 
To take into account fixed points, for which $\sigma=0$, we introduce the number of states with a given coupling constant $J$, $\mathcal{N}(J;T) \sim \exp[NS(J;T)]$.
Along the same lines as in the main text, we find
\be 
S(J; T) = \ln 2-\frac{1}{\xi^2} \big(J-J_0\big)^2-\frac{U^2}{4T^2}
\ee
if $|J-J_0|<\xi\sqrt{\ln2-U^2/(4T^2)}$ and $S(J; T)=0$ otherwise. 
The configurational entropy of the fixed points is defined as
\be S_{\mathrm{fp}}^*(T, \mu)=\underset{{\sigma(J; T,\mu)=0}}{\max} \,S(J; T).\ee
We now discuss the value of $S_{\mathrm{fp}}^*(T, \mu)$ for $\mu>\mu_0(T)$. According to the main text, $\sigma(J; T, \mu)=0$ for all $T\geq J/2$. We thus find that for $T\geq T_0$, the maximum is reached for $J=J_0$ which leads to
\be S_{\mathrm{fp}}^*(T, \mu)=\ln 2-\frac{U^2}{4T^2}.\ee
For $T<T_0$, the maximum of $S(J; T)$ under the constraint $\sigma(J;T, \mu)=0$ is reached for $J=2T$, leading to 
\be S_{\mathrm{fp}}^*(T, \mu)=\ln 2 -\frac{(2T-J_0)^2}{\xi^2}-\frac{U^2}{4T^2} \ee
if $\ln 2 -(2T-J_0)^2/\xi^2-U^2/(4T^2)>0$ and 
$ S_{\mathrm{fp}}^*(T, \mu)=0$ 
otherwise.
$S_{\mathrm{fp}}^*(T, \mu)$ is plotted in Fig.~2 of the main text.

\end{document}